\documentclass[journal]{IEEEtran}
\usepackage{cite}
\usepackage[pdftex]{graphicx}
\usepackage{caption}
\usepackage{subcaption}
\usepackage{amsmath}
\usepackage{enumerate}
\usepackage{array}
\usepackage{float}
\usepackage{xcolor}

\begin{document}
 \title{Towards Relevance and Sequence Modeling in Language Recognition}
 \author{Bharat~Padi, Anand Mohan and
         Sriram~Ganapathy,~\IEEEmembership{Senior Member,~IEEE}
  \thanks {This work was partly supported by grants from Department of Science and Technology, Early Career Award (ECR-01341/2017) and the Extra Mural Grant (EMR-2016/007934).}
  \thanks{Bharat Padi is with the mind.ai, Bangalore. This work was performed when he was a student at the LEAP lab. }
  \thanks{
  Anand Mohan is with Amazon Alexa Lab, Bangalore, India. This work was performed when he was a student at the LEAP lab.}
  \thanks{
  Sriram Ganapathy is with the Learning and Extraction of Acoustic Patterns (LEAP) lab, Electrical Engineering, Indian Institute of Science, Bengaluru, India, 560012}}

\maketitle

\begin{abstract}
The task of automatic language identification (LID) involving multiple dialects of the same language family in the presence of 
noise   is a challenging problem. In these scenarios, the identity of the language/dialect may be reliably present only in parts of the temporal sequence of the speech signal. The conventional approaches to LID (and for speaker recognition) ignore the sequence information by extracting long-term statistical summary of the recording assuming an independence of the feature frames. In this paper, we propose a neural network framework utilizing short-sequence information in language recognition. In particular, a new model is proposed for incorporating relevance  in language recognition, where parts of speech data are weighted more based on their relevance for the language recognition task. This relevance weighting is achieved using the bidirectional long short-term memory (BLSTM) network with attention modeling. We explore two approaches, the first approach uses segment level i-vector/x-vector representations that are aggregated in the neural model and the second approach where the acoustic features are directly modeled in an end-to-end neural model.  Experiments are performed using the language recognition task in NIST LRE 2017 Challenge using clean, noisy and multi-speaker speech data as well as in the RATS language recognition corpus. In these experiments on noisy LRE tasks as well as the RATS dataset, the proposed approach yields significant improvements over the conventional i-vector/x-vector based language recognition approaches as well as with other previous models incorporating sequence information.  
\end{abstract}


\IEEEpeerreviewmaketitle

\section{Introduction}

\IEEEPARstart{T}{he} problem of recognizing the spoken language of a given audio segment is 
of considerable interest for several commercial applications like speech translation \cite{waibel}, multi-lingual speech recognition \cite{schultz2001language}, document retrieval \cite{chelba2008retrieval} as well as in defense and surveillance applications \cite{walker2012rats}. In the recent years, several advances in speech signal modeling, most prominent of them being the application of factor analysis methods, have contributed to improving the performance of language recognition systems \cite{li2013spoken}. However, the task is still of considerable challenge when the recognition involves multiple dialects of the same language (like the recent NIST Language Recognition Evaluation (LRE) 2017 challenge). \textcolor{black}{Further, the language identification (LID) performance is degraded  in the presence of noise and other artifacts, such as in the robust automatic transcription of speech (RATS) databases \cite{walker2012rats,  ganapathy2013unsupervised, mclaren2016analyzing}.  In this paper, we propose a modeling framework to address the challenge of noise in language recognition. }

Traditionally, phoneme recognition followed by language modeling (PRLM) was one of the popular methods for automatic LID task~\cite{zissman,navratil}. 
In the recent past, the use of deep neural network (DNN) based posterior features were attempted for LID \cite{benzeghiba}. The bottleneck features based on the acoustic model of a speech recognition system had also shown promising results for noisy language
 recognition \cite{richardson2015deep}. 

The development of i-vectors as one of the primary representations for LID was first introduced in \cite{dehak2011language}. The i-vectors are features of fixed dimension derived from variable length speech utterances using a background model \cite{dehak2011front}. The background model can be a Gaussian mixture model (GMM) \cite{reynolds2000speaker} or a DNN model \cite{richardson2015deep,lei2014novel}. The i-vectors extracted from the training data are used to train classifiers such as support vector machines (SVMs) which perform the task of language identification \cite{ganapathy2014robust,padi2018leap}. 

One of the main drawbacks of the i-vector representations \cite{dehak2011language} and the recently proposed x-vector representations \cite{snyder2015time} is the long term summarization of the audio signal.  \textcolor{black}{Even in the extraction of x-vector embeddings using the TDNN models~\cite{snyder2018spoken}, the temporal context of $15$ frames (about 150msec.) is alone used in the forward propagation of frame level features. }
For tasks like dialect identification in the presence of noise and other artifacts, some regions of audio may be more reliable than the rest.   
We hypothesize that there is a need to extract information only from the relevant regions of the audio signal rather than the long-term summary of the signal for the task of noisy LID.  The attention approach in neural network modeling was originally proposed for neural machine translation \cite{bahdanau2014neural} and image captioning \cite{xu2015show}. 
The attention approach to speech recognition was initially investigated for phoneme recognition \cite{chorowski2015attention} where the issues in dealing with variable length speech sequences were identified. Recently, the attention based models have also been applied for end-to-end speech recognition tasks \cite{bahdanau2016end,watanabe2017hybrid,bansal2019speaker}.  
The end-to-end approaches to language recognition have been explored with long short term memory (LSTM) networks and with DNNs \cite{mounika2016investigation}. A recent approach using curriculum learning had also been applied for noise robust language recognition \cite{vuddagiri2018curriculum}. 
However, the state-of-the-art language recognition systems using large scale NIST language recognition evaluation (LRE) challenges, continue to use the i-vector/x-vector based approaches with support vector machine classifier \cite{nistlre17_odyssey18}. 

\textcolor{black}{In this paper, we propose an attention based framework for language recognition to perform relevance weighting  of the audio signal region in the presence of noise.}
 The term relevance used in this work corresponds to the relative importance of short regions of audio ($1000$msec. chunks) for determining the language/dialect of the given utterance.  
 For neural modeling of relevance, we explore two approaches,
\begin{enumerate}[i]
\item Modeling short segment statistics from i-vector representations using a deep bidirectional LSTM model~\cite{padi2019attention}. We refer to this as the i/x$-$BLSTM model.
\item Modeling the audio features directly using end-to-end deep recurrent model with hierarchical gated recurrent units. We refer to this model as HGRU \cite{padi2019icassp}.
\end{enumerate}


The rest of the paper is organized as follows. The state-of-the-art systems using i-vector based model  and the LSTM based neural network approach to end-to-end language recognition are discussed in Sec.~\ref{sec:ivec_baseline} and Sec.~\ref{sec:lstm_baseline} respectively. Then, we discuss several approaches for incorporating relevance using short-term i-vector representations in Sec.~\ref{sec:relevance}. In Sec.~\ref{sec:hybrid_ivecLSTM}, we discuss the proposed hybrid model of using short-term i-vector representations in a neural framework for language recognition. Sec.~\ref{sec:hgru} details the proposed end-to-end framework for language recognition. The details of the experimental set and the various datasets used is given in Sec.~\ref{sec:expt}. The LID results of various language recognition tasks on LRE 2017 and RATS datasets are reported in Sec.~\ref{sec:results}.  This is followed by an analysis of the robustness of the proposed approaches in the presence of noise 
in Sec.\ref{sec:discussion}. In Sec.~\ref{sec:summary}, we summarize the important contributions of this paper.

\section{Relevant Prior Work}


\subsection{i-vector/x-vector SVM LID system}\label{sec:ivec_baseline}
The i-vectors constitute the widely used features for language recognition~\cite{dehak2011language}. 
A Gaussian Mixture Universal Background Model (GMM-UBM) is obtained by pooling the front end features from all the utterances in the train dataset. The means of the GMM are adapted to each utterance using the Baum-Welch (BW) statistics of the front-end features. 
A Total Variability Model (TVM) is assumed as a generative model for the adapted GMM mean supervector, which is given by,
    \begin{align*}
	\boldsymbol{M}(s) &= \boldsymbol{M}_0 + T\boldsymbol{y}(s),
\end{align*}
where $ \boldsymbol{M}_0$ and $ \boldsymbol{M}(s)$ are the UBM mean supervector and the adapted mean supervector of recording $s$ respectively. Here, $T$ is a rectangular matrix, and $\boldsymbol{y}(s) \sim \mathcal{N}(\boldsymbol{0},I)$ is a latent variable. 
The \textit{maximum aposteriori} (MAP) estimate of $\boldsymbol{y}(s)$ given the front-end features of the recording $s$ is called the i-vector of recording $\boldsymbol{y}^*(s)$. The i-vectors extracted for each of the speech files are processed with length normalization \cite{garcia2011analysis} and dimensionality reduction is  performed using  linear discriminant analysis (LDA). 

\textcolor{black}{Recently, x-vector representations have been developed for speaker and language recognition~\cite{snyder2015time,snyder2018spoken}. The x-vector model is based on time delay neural network (TDNN) where the initial layers operate on frame level features. The higher layers convert these features to segment level by summarizing mean and standard deviation of frame level features. This is followed by $2$ fully connected hidden layers and the entire neural network model is trained to classify languages. The utterance level features before the last fully connected hidden layer are used as embeddings and they are termed as x-vectors.  The x-vector features can be used instead of i-vector embeddings in a SVM classifier for language recognition~\cite{snyder2018spoken}.}

\subsection{Long Short Term Memory Neural Network LID system}\label{sec:lstm_baseline}

Long Short-Term Memory networks (LSTM) \cite{hochreiter1997long,gers1999learning} are designed to address the issue of learning long-term dependencies in recurrent neural networks \cite{hochreiter2001gradient}.  
In \cite{lopez2014automatic,2016_odyssey_lstm_nist_Zazo}, the authors have proposed an Long Short Term Memory Recurrent Neural Network (LSTM-RNN) based end-to-end model for exploiting temporal information useful for LID. In \cite{lopez2014automatic} authors have shown that the LSTM model outperforms a baseline i-vector system and a deep neural network model on short duration ($3$s) test segments. In a recent work, the authors of \cite{2016_odyssey_lstm_nist_Zazo} show that neural networks are useful even for challenging NIST datasets where the LSTM model outperforms i-vector baseline on short duration ($3$ sec.) test segments but performs relatively worse on longer duration test segments ($10$ sec., $30$ sec.). 


\section{Relevance In Language Recognition}\label{sec:relevance}
We propose three different approaches for incorporating relevance and sequence information in language recognition.

\subsection{Relevance Weighted Baum-Welch statistics}\label{sec:rwbw}

The main motivation in this approach is the use of inverse entropy as measure of confidence/relevance. \textcolor{black}{The entropy of class posterior distribution carriers information regarding the  uncertainty of the classification. Given a discrete class posterior distribution, the higher the entropy the lower the confidence of the decision.}  
Misra et. al~\cite{misra2003new} had shown that the entropy increases with decrease in SNR as the posterior probability distribution becomes more and more uniform when the noise level increases. If two segments $s1,s2$ of the audio contain two different SNR values and if the corresponding i-vector representations be denoted as $y_{1}$ and $y_{2}$, the inverse of the entropy of the corresponding language posteriors $H(l|y_{1})$ and $H(l|y_{2})$ provides a measure of the reliability of the decisions from each of audio regions. The optimal language decision using the information from the two audio regions can be taken by using inverse entropy weighted linear combination of the posteriors ~\cite{misra2003new,valente2009novel},
\begin{eqnarray}
\color{black}{l^* = argmax_{l} \big[ w_1p(l|y_{1}) + w_2p(l|y_{2})} \big] 
\end{eqnarray}
\textcolor{black}{where, $w_1$ and $w_2$ are weights given by}
\begin{eqnarray}
\color{black}{w_1 = \frac{\frac{1}{H(l|y_{1})}}{\frac{1}{H(l|y_{1})}+\frac{1}{H(l|y_{2})}}}
\end{eqnarray}
\textcolor{black}{and $w_2 = 1 - w_1$. In the proposed work, the relevance weighting of the Baum-Welch statistics is used to provide the optimal statistics for i-vector estimation (similar to SAD probability estimation done in~\cite{mclaren2015softsad}).}
\begin{figure}[t!]
\centering
\includegraphics[width=8.4cm]{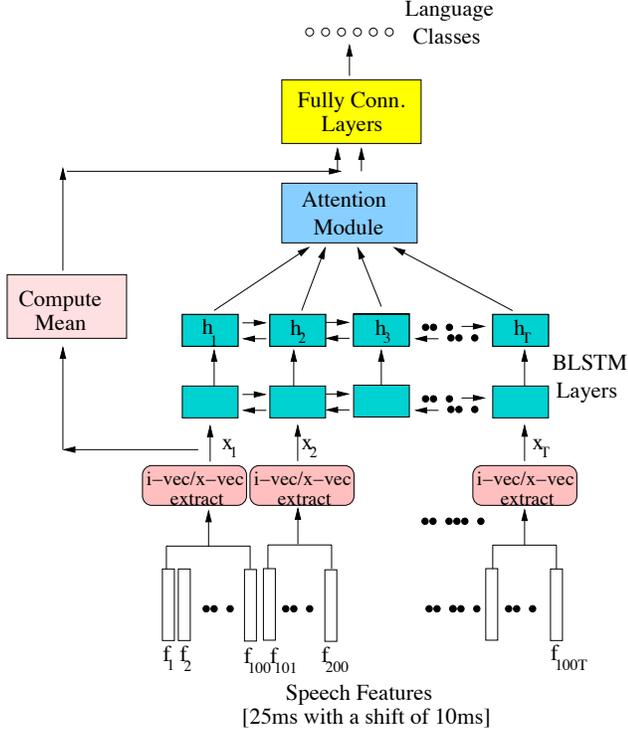}
\caption{Proposed BLSTM model using $1000$ msec. i-vector/x-vector features for language recognition.}
\label{fig:blstm}
\vspace{-0.15 in}
\end{figure}

In our proposed approach, we use short term i-vectors that are labeled individually according to the label of the corresponding utterance. These $1000$ msec. i-vectors from train and development datasets are then used to train a feed forward deep neural network (DNN). The DNN has an input layer of $500$ dimensions with $3$ hidden layers having $1024$ dimensions and rectified linear unit (ReLU) non-linearity. The output layer was $L$ dimensions where $L$ is the number of language classes. 

Once the DNN model is trained, for each non-overlapping short term i-vector $\boldsymbol{y}_i$ from every utterance, the information entropy $H_i$ is computed from the posteriors from the DNN model 
where $i$ is the index of $1000$ msec. i-vector. 
We use the entropy measure $H_i$ to compute a relevance parameter $\gamma_i$. 
The value of $\gamma_i$ changes only at $1$ sec. intervals.
The value of $\gamma_i$ is obtained from $H_i$ as,
\begin{align}
\gamma_i &= \begin{cases}
1 & \text{when $H_i < H_{\mathsf{min}}$}\\
\frac{H_{\mathsf{max}}-H_i}{H_{\mathsf{max}}-H_{\mathsf{min}}} & \text{when $H_{\mathsf{min}} \leq H_i \leq H_{\mathsf{max}} $}\\
0 & \text{when $H_i > H_{\mathsf{max}}$}
\end{cases}
\end{align}

where $H_{\mathsf{max}}$ and $H_{\mathsf{min}}$ are hyper-parameters.
\begin{figure}[t!]
\centering
\includegraphics[width=8.5cm,scale=1.5]{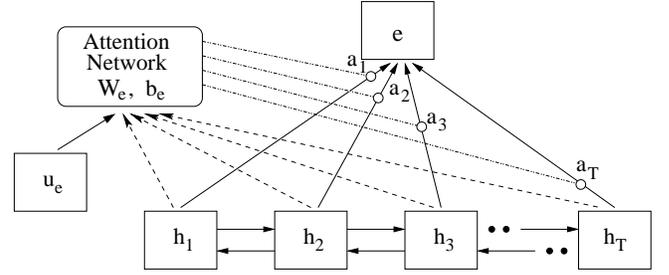}
\caption{ Attention modeling in the proposed i/x-BLSTM/HGRU model.}
\label{fig:attn}
\end{figure}
The zeroth and first-order Baum-Welch statistics used for i-vector extraction are now modified as:
\begin{align}
N_c(s) &= \sum_{i=1}^{H(s)}\gamma_i \operatorname{p}(c\,|\,\boldsymbol{x}_i) \\
	\boldsymbol{F}_{X,c}(s) &= \sum_{i=1}^{H(s)}\gamma_i \operatorname{p}(c\text{ }|\text{ }\boldsymbol{x}_i)(\boldsymbol{x}_i - \boldsymbol{\mu}_c)
\end{align}

\subsection{Hybrid i-vector/x-vector BLSTM model}\label{sec:hybrid_ivecLSTM}
This approach is shown in Fig.~\ref{fig:blstm}. 
The short-term i-vectors/x-vectors, extracted every $200$ msec. from overlapping windows of $1000$ msec. duration ($100$ frames of acoustic features $f_1, f_2,...$ extracted with $10$ msec. shift) are modeled using a bidirectional LSTM model (BLSTM) with attention. 
The input to the BLSTM is the variable length sequence of  short-term i-vectors/x-vectors. The LSTM architecture that we use in this paper contains two layers with $256$ memory cells in each layer\footnote{All models proposed in this work are available at https://github.com/iiscleap/lre-relevance-weighting}.
 We propose to use an attention model to weight the $1$ sec. representations based on their relevance to the language classification task. The attention method \cite{bahdanau2014neural} provides an efficient way to aggregate the sequence of $1$ sec. vectors. The attention mechanism used in this work is shown in Fig.~\ref{fig:attn}. The model implements the following set of equations,
\begin{align}
\textbf{u}_{t} &= tanh(\textbf{W}_e\textbf{h}_{t} +\textbf{ b}_e)\\
{a}_{t} &= \frac{exp(\textbf{u}_{t}^T\textbf{u}_e)}{\sum_t{exp(\textbf{u}_{t}^T\textbf{u}_e)}}\\\label{eq:eq_attn}
\textbf{{e}} &= \sum_t{a_{t}\textbf{h}_{t}}
\end{align}
Here, $\textbf{W}_e$, $\textbf{b}_e$ are the weights and the bias of the attention module which are learned in training process along with the vector $\textbf{u}_e$.  The output  $\textbf{{e}}$ denotes the fixed dimensional embedding from the input sequence. The attention module based on the similarity of $\textbf{u}_t$ with $\textbf{u}_e$ assigns normalized weights $\textbf{a}_t$ using a softmax function. 
 The utterance level representation $\textbf{{e}}$ is  mapped to the final language targets through a layer of fully connected network (FCN) with  $512$ dimensions  and ReLU non-linearity. The output layer has a softmax non-linearity and the model is trained with cross entropy loss  using stochastic gradient descent  \cite{werbos1990backpropagation} 
and is implemented in TensorFlow \cite{abadi2016tensorflow}.

\begin{figure}[t!]
\centering
\includegraphics[width=\columnwidth]{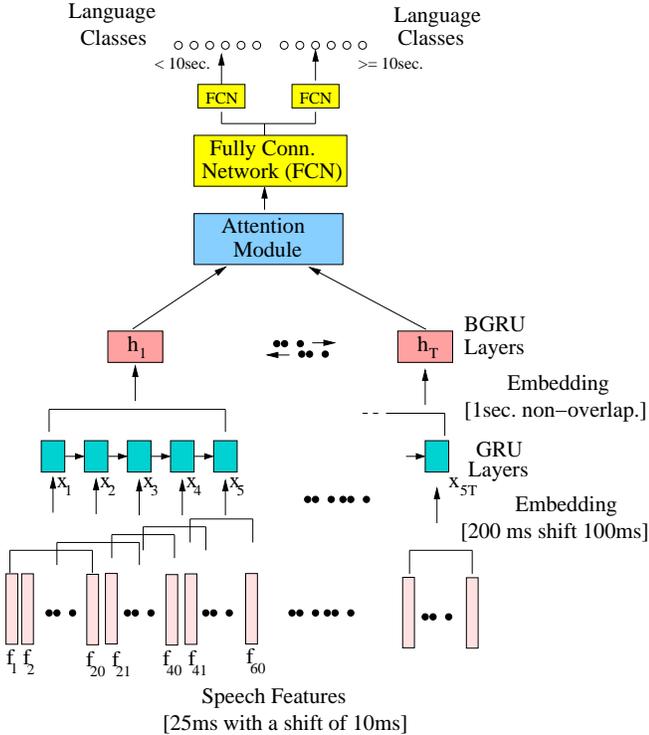}
\caption{End-to-end Hierarchical GRU RNN with attention module and duration dependent target layers.}
\label{fig:grurnn}
\end{figure}
\subsection{End-to-End Hierarchical GRU model}\label{sec:hgru}
We propose to use a simplified version of the LSTM function, the Gated Recurrent Unit (GRU) architecture \cite{cho2014learning} in the end-to-end model. The GRUs combine the input and forget gate into one single gate and in many tasks, the GRUs that have a relatively smaller number of parameters are shown to achieve or improve over the performance of LSTMs \cite{69e088c8129341ac89810907fe6b1bfe}. Even with LSTM/GRU models, the direct modeling of long sequences can be cumbersome~\cite{2016_odyssey_lstm_nist_Zazo}. In order to model such long sequences, we propose a novel hierarchical bidirectional GRU \cite{padi2019icassp} network with attention in this paper.





The block schematic of the proposed model is given in Fig.~\ref{fig:grurnn}. 
At the first layer, a $256$ cell unidirectional GRU block accumulates information across a window of $200$ msec. i.e, a sequence of $20$ feature vectors with a shift of $100$ msec. over the entire input sequence. The output from the first layer is a sequence of vectors that are sampled every $100$ msec. with each vector representing information from overlapping $200$ msec. segments of input speech. This is then fed to the second layer of GRU block with $512$ cells where the information is accumulated over a window of $1$ sec.   The accumulated $1$ sec. vectors from second layer are fed to the final bidirectional GRU layer \cite{Schuster97bidirectionalrecurrent} with $512$ cells in the third layer.

\begin{table}[t!]
\caption{\label{table:lre-languages} {LRE17 training set : language clusters, target languages and total number of hours.}}
\vspace{2mm}
\centerline{
\begin{tabular}{l l l}
\hline
Cluster & Target Languages & Hours \\ \hline
Arabic & \begin{tabular}[l]{@{}l@{}}Egyptian Arabic (ara-arz)\\ Iraqi Arabic (ara-acm)\\ Levantine Arabic (ara-apc)\\ Maghrebi Arabic (ara-ary)\end{tabular} & \begin{tabular}[l]{@{}l@{}}190.9\\130.8\\440.7\\81.8\end{tabular}\\\hline
Chinese& \begin{tabular}[l]{@{}l@{}}Mandarin (zho-cmn)\\ Min Nan (zho-nan)\end{tabular} & \begin{tabular}[l]{@{}l@{}}379.4\\13.3\end{tabular}\\\hline
English &\begin{tabular}[l]{@{}l@{}}British English (eng-gbr)\\ General American English (eng-usg)\end{tabular} & \begin{tabular}[l]{@{}l@{}}4.8\\327.7\end{tabular}\\\hline
Slavic & \begin{tabular}[l]{@{}l@{}} Polish (qsl-pol)\\ Russian (qsl-rus)\end{tabular} & \begin{tabular}[l]{@{}l@{}}59.3\\69.5\end{tabular}\\\hline
Iberian& \begin{tabular}[l]{@{}l@{}}Caribbean Spanish (spa-car)\\ European Spanish (spa-eur)\\ Latin American Continental Spanish (spa-lac)\\Brazilian Portuguese (por-brz)\end{tabular} & \begin{tabular}[l]{@{}l@{}}166.3\\24.7\\175.9\\4.1\end{tabular}\\\hline
\end{tabular}}
\end{table}

The output of the three layer hierarchical GRU model contains representations at $1$ sec. level.  The rest of model framework is similar to the $i$-$BLSTM$ model (Sec.~\ref{sec:hybrid_ivecLSTM}). 
For the end-to-end model, we found that the distribution of the embeddings from the attention network is quite different for short and long duration inputs.
Hence, we propose to use two separate target layers, one for short duration inputs that are of the order of $3$ sec. duration and other one with longer duration input sequences that are $10$ sec. or above. 
The entire network is trained using Adam optimization and Back Propagation Through Time (BPTT) algorithm \cite{werbos1990backpropagation}.

\begin{table*}[t!]
\centering
\caption{Performance of reference and the developed systems on the NIST LRE17 evaluation dataset in terms of percentage accuracy, $C_{avg}$ and EER. \textcolor{black}{The performance of the x-vector systems are shown in parentheses. The x-vector systems were evaluated only for $3$,$10$ and $30$s conditions. The x-post system uses the x-vector network directly without the proposed BLSTM backend model.}}
\label{table:lre17_perf}
\begin{tabular}{|c|c|c|c||c|c|c|c|c|}
\hline

Dur. & i-LDA-SVM \textcolor{black}{(x-LDA-SVM)} & LSTM \cite{2016_odyssey_lstm_nist_Zazo}  & DNN-Attn \cite{vuddagiri2018curriculum}  & RWBW &  i-BLSTM \textcolor{black}{(x-BLSTM)} & HGRU & \textcolor{black}{x-post.} & \textcolor{black}{x-BLSTM-E2E} \\ \hline \hline\multicolumn{9}{|c|}{ Accuracy (\%)}                                      \\ \hline \hline 
3               & 53.84 \textcolor{black}{(46.14)}   & 54.7  & 54.6     & 51.37      & 54.80 \textcolor{black}{(56.99)}   & {55.13} & \textcolor{black}{({60.55})}  & \textcolor{black}{(\textbf{60.2})}\\ 
10              & 72.36 \textcolor{black}{(73.19)}  & 72.1  & 72.5     & 68.42     & \textbf{75.89} \textcolor{black}{(71.60)}   & 74.06 & \textcolor{black}{({69.82})} &  \textcolor{black}{(70.5)}\\ 
30              & {82.98} \textcolor{black}{(\textbf{85.09})}  & 76.1  & 79.7     & 77.59     & 82.27  \textcolor{black}{(76.02)}  & {82.98}  & \textcolor{black}{({72.02})} &  \textcolor{black}{(74.2)} \\ 
1000            & 56.23   & 42.8  & 50.2     & \textbf{58.48}     & 54.07    & 53.53 & - & -\\ \hline
overall         & 67.86   & 64.3  & 66.36    & 64.78     & \textbf{68.65}    & 68.48 & - & -\\ \hline \hline
\multicolumn{9}{|c|}{$C_{avg}$}                                             \\ \hline \hline

3               & 0.53  \textcolor{black}{(0.49)}  & 0.55  & 0.53     & 0.58     & {0.50} \textcolor{black}{({0.46})}    & 0.55  & \textcolor{black}{(\textbf{0.45})} & \textcolor{black}{({0.46})} \\ 
10              & 0.27 \textcolor{black}{(\textbf{0.21})}   & 0.35  & 0.28     & 0.35     & {0.26}  \textcolor{black}{(0.28)}   & 0.32 &\textcolor{black}{(0.32)} &\textcolor{black}{(0.32)} \\ 
30              & {0.13} \textcolor{black}{(\textbf{0.10})}    & 0.28  & 0.20     & 0.21      & 0.18 \textcolor{black}{(0.23)}    & 0.23  & \textcolor{black}{(0.28)} & \textcolor{black}{(0.28)}\\ 
1000            & 0.54    & 0.79  & 0.61     & 0.51  & \textbf{0.50}     & 0.62 & - & -\\ \hline
overall         & 0.37    & 0.48  & 0.40     & 0.40     & \textbf{0.36}     & 0.42 & - & -\\ \hline \hline
\multicolumn{9}{|c|}{ EER (\%)}                  \\ \hline
\hline
3               & \textbf{13.40} \textcolor{black}{(16.92)}   & 15.39 & 14.98    & 15.60      & 15.47 \textcolor{black}{(13.50)}   & 15.33 &\textcolor{black}{(13.70)} & \textcolor{black}{(13.45)}\\ 
10              & 6.47 \textcolor{black}{(\textbf{5.98})}   & 8.70  & 7.05     & 7.84     & {6.32}  \textcolor{black}{(7.33)}   & 7.49  &\textcolor{black}{(9.20)} &  \textcolor{black}{(8.53)} \\ 
30              & {3.50} \textcolor{black}{(\textbf{2.75})}   & 7.25  & 4.73     & 4.61     & 3.67  \textcolor{black}{(6.28)}   & 4.93 & \textcolor{black}{(8.11)} & \textcolor{black}{(7.12)} \\ 
1000            & 15.35   & 26.27 & 15.72    & \textbf{14.20}     & 14.71    & 17.02 & - & -\\ \hline
overall         & \textbf{9.26}    & 14.38 & 10.80    & 10.34     & 9.65    & 10.79 & - & - \\ \hline
\end{tabular}
\end{table*}

\section{Experimental set up} \label{sec:expt}

\subsection{Feature Extraction} All the systems in this paper use the same front end features. The features are the Deep Neural Network (DNN) Bottleneck (BN) features \cite{richardson2015deep,nistlre17_odyssey18}.  We extract BN features from a feed forward deep neural network trained for automatic speech recognition (ASR) using Kaldi \cite{povey2011kaldi} framework. The ASR-DNN was trained using $39$ ($13 + \varDelta + \varDelta\varDelta$) dimensional mel frequency cepstral coefficient (MFCC) features with $10$ msec. frame shift and  $25$ msec. windows. The ASR-DNN was trained on speech from combined Switchboard (SWB1) and Fisher corpora (about $2000$ hours of labeled audio). The ASR-DNN model uses $7$ hidden layers with ReLU activation with layer-wise batch normalization. We use the last hidden layer output of size $80$. 
\subsection{Datasets}
\subsubsection{NIST LRE 2017 Dataset}

The details of the LRE17 dataset is given in Table~\ref{table:lre-languages}. The LRE17 training data (LDC2017E22) has five major language clusters with 14 target dialects  with a total duration of $2069$ hours in $16205$ files. The development dataset consists of $3661$ files which contain $253$ hours of audio and the evaluation dataset consists of $25451$ files with $1065$ hours of audio. The development and evaluation datasets are further partitioned into utterances of durations $3$ sec, $10$ sec, $30$ sec. or $1000$ sec. The datasets contain conversational telephone speech (CTS) and broadcast narrow band speech (BNBS) and speech extracted from videos or video speech (VS) (the $1000$ sec. files). 
\textcolor{black}{We have not used the development data in training the embedding models (i-vector/x-vector) or the back-end models.}

In addition to the standard LRE17 test set, we test the robustness of the various LID systems using noisy versions of the evaluation data for the $10$ sec. recording conditions. We use four different noise types (Babble, Subway, Airport and Street) from the Aurora-4 corpus and these are added to the audio signal at different signal-to-noise ratio (SNR) like $5$, $10$, $15$ and $20$ dB using the filter and noise adding tool (FANT) \cite{hirsch2005fant}. While adding the noise, either the entire audio file is corrupted with noise ($Noisy$ cond.) or only the first half of the audio file is corrupted with noise ($Partial~ Noise.$). The latter condition is considered to simulate non-stationary test conditions where the noise characteristics can change within the course of recording duration. 

\begin{table*}[t!]
\centering
\caption{Performance of the systems when evaluated on data ($10$ sec. recording condition) corrupted with noise at various SNR levels. \textcolor{black}{The performance of the x-vector systems are shown in parentheses.} }
\label{table:noisy}
\begin{tabular}{|c|c|c|c|c|c|c|c|}
\hline

SNR & i-LDA-SVM \textcolor{black}{(x-LDA-SVM)} & LSTM \cite{2016_odyssey_lstm_nist_Zazo} & DNN-Attn \cite{vuddagiri2018curriculum} &  i-BLSTM \textcolor{black}{(x-BLSTM)} & HGRU & \textcolor{black}{(x-post)}  & \textcolor{black}{(x-BLSTM-E2E)} \\ \hline
\multicolumn{7}{|c|}{Accuracy (\%)}                           \\ \hline
5dB       & 47.28 \textcolor{black}{(49.96)}  & 48.36 & 46.35 & {51.58} \textcolor{black}{(55.44)}  &  45.78 & \textcolor{black}{(\textbf{58.13})}
& \textcolor{black}{({57.41})}\\ 
10dB      & 53.42 \textcolor{black}{(58.43)}  & 56.30 & 54.86 & {59.86} \textcolor{black}{(62.65)}   &  53.92 & \textcolor{black}{({63.57})} & \textcolor{black}{(\textbf{63.73})}\\ 
15dB      & 57.47 \textcolor{black}{(63.20)}  & 61.63 & 61.58 & {64.30} \textcolor{black}{(66.27)}  &   60.20 & \textcolor{black}{({65.88})}  &  \textcolor{black}{(\textbf{66.94})}\\ 
20dB      & 59.64 \textcolor{black}{(66.66)}  & 65.28 & 65.76 & {67.72} \textcolor{black}{(68.24)} &   64.21 & \textcolor{black}{({67.40})} & \textcolor{black}{(\textbf{68.36})}\\ \hline
Avg. & 54.45 \textcolor{black}{(59.56)} & 57.89  & 57.13 & {60.87} \textcolor{black}{(63.15)} &  56.02 & \textcolor{black}{({63.74})}  &  \textcolor{black}{(\textbf{64.36})}\\ \hline \hline 
\multicolumn{7}{|c|}{Cavg}                           \\ \hline
5dB       & 0.62 \textcolor{black}{(0.71)} & 0.65 & 0.60 & {0.59} \textcolor{black}{(\textbf{0.47})} &  0.68 & \textcolor{black}{({0.47})} & \textcolor{black}{(0.50)}\\ 
10dB      & 0.53 \textcolor{black}{(0.63)}& 0.54 & 0.49 & {0.48} \textcolor{black}{(\textbf{0.38})} & 0.58 & \textcolor{black}{({0.40})} &   \textcolor{black}{(0.41)} \\ 
15dB      & 0.48 \textcolor{black}{(0.58)} & 0.47 & {0.41} & 0.42 \textcolor{black}{(\textbf{0.33})} &  0.48 & \textcolor{black}{({0.36})} & \textcolor{black}{(0.37)} \\ 
20dB      & 0.45 \textcolor{black}{(0.54)} & 0.42 & {0.37} & {0.37} \textcolor{black}{(\textbf{0.31})} &  0.43 & \textcolor{black}{({0.34})}& \textcolor{black}{(0.35)} \\ \hline
Avg. & 0.52 \textcolor{black}{(0.62)} & 0.52 & {0.47} & {0.47} \textcolor{black}{(\textbf{0.37})} &  0.54 & \textcolor{black}{({0.39})} & \textcolor{black}{(0.41)} \\ \hline \hline
\multicolumn{7}{|c|}{EER (\%)}                           \\ \hline
5dB       & 22.78 \textcolor{black}{(26.52)} & 19.79 & 17.95 & {17.01} \textcolor{black}{(\textbf{14.44})} & 19.16 & \textcolor{black}{({16.25})}& \textcolor{black}{(15.54)}  \\ 
10dB      & 18.20 \textcolor{black}{(21.97)} & 15.88 & 13.27 & {12.90} \textcolor{black}{(\textbf{11.11})} & 15.61 & \textcolor{black}{({12.64})}& \textcolor{black}{(11.94)} \\ 
15dB      & 15.63 \textcolor{black}{(19.30)} & 12.87 & {10.88} & 11.05 \textcolor{black}{(\textbf{9.26})} & 13.19  & \textcolor{black}{({10.95})}& \textcolor{black}{(10.07)}\\ 
20dB      & 13.99 \textcolor{black}{(17.87)} & 11.11 & {9.43} & 9.44 \textcolor{black}{(\textbf{8.37})}  &  11.69  & \textcolor{black}{({10.11})}& \textcolor{black}{(9.20)}\\ \hline
Avg. & 17.65 \textcolor{black}{(21.41)} & 14.91 & 12.88 & {12.60} \textcolor{black}{(\textbf{10.80})}  & 14.91  & \textcolor{black}{(12.49)} & \textcolor{black}{(11.69)} \\ \hline 
\end{tabular}
\end{table*}

\subsubsection{RATS LID Dataset} 
The DARPA Robust Automatic Transcription of Speech (RATS)~\cite{walker2012rats} program targets the development of speech systems operating on highly distorted speech recorded over ``degraded"  radio channels.
The data used here consists of recordings obtained from re-transmitting a clean signal over eight different radio channel types, where each channel introduces a unique degradation mode specific to the device and modulation characteristics~\cite{walker2012rats}. For the language identification (LID) task, the performance is degraded due to the short segment duration of the speech recordings in addition to the significant amount of channel noise \cite{ibm13}.

The training data for the RATS experiments consist of $20000$ recordings (about $1600$ hours of audio) from five target languages (Arabic, Pashto, Dari, Farsi and Urdu) as well as from several other non-target languages. We have used 6 out of 8 given channels (channels B-G) for training and testing purposes. The development and the evaluation data consists of $5663$ and $14757$ recordings respectively from the above 6 channels. We also evaluate the models on sampled $3$ sec, $10$ sec. and $30$ sec. chunks of voiced data from the full length evaluation files.

\subsection{Evaluation Metrics}
The performance is measured using the primary metric described in the evaluation plan of NIST LRE 2017 $C_{avg}$ ~\cite{nistlre17_odyssey18}, Equal Error Rate (EER) and accuracy. 
\begin{itemize}
\item $\mathbf{C_{avg}}$ - 
The pair-wise language recognition performance will be computed for all the target-language/non-target-language pairs $(L_T, L_N)$. An average performance cost for each system is computed as
\begin{align}\nonumber
C_{avg}(\beta) &=  \frac{1}{N_L}\sum_{L_T} \left\{P_{m}(L_T) + \sum_{L_N} \frac{\beta  P_{f}(L_T, L_N)}{N_L-1}\right\}
\end{align}
where $P_{m}$ is the probability of miss detection, $P_{f}$ is the probability of false alarm, which are computed by applying detection threshold of $\log{\beta}$ to  the system scores. The primary metric used for LRE17 is the average cost of two $C_{avg}$ measures obtained using $\beta_1 = 1$, $\beta_2 = 9$. 

\item \textbf{EER} - Equal Error Rate is measured for individual languages and then their average is computed. 
Note that, the $EER$ and $C_{avg}$ consider the LID system as detection problem.

\item \textbf{Accuracy} - The accuracy is measure in a classification setting. It considers the LID systems as multi-class learning problem ($14$ classes in LRE17 dataset and $6$ classes in the RATS evaluation). 
\end{itemize}
\subsection{i-vector/x-vector baseline system}
Once the BN features are extracted from the ASR-DNN for the LID data, a  speech activity detection (SAD) algorithm was applied to remove the unvoiced frames~\cite{SAD}. We use the implementation of SAD from the Voicebox toolkit \cite{brummer}. This was followed by cepstral mean variance normalization (CMVN) done over each utterance, followed by a sliding window CMVN over a sliding window of $3$ sec. The number of UBM mixture components was set to $C=2048$ and the dimension of the total variability space was fixed to be $R=500$. 
The i-vectors  are processed with within class covariance normalization (WCCN) technique~\cite{hatch2006within} and length normalized. The dimension of the i-vectors is then reduced to $13$ and $5$ for LRE17 and RATS datasets respectively using linear discriminant analysis (LDA) and  modeled using support vector machines (LDA-SVM). The i-vector SVM system implementation follows the recipe provided by the NIST LRE system \cite{nistlre17_odyssey18}. 

\textcolor{black}{The x-vector baseline system is implemented using the TDNN model architecture described in \cite{snyder2018spoken}. The x-vector system consists of $5$ layers of time delay neural network (TDNN) architecture followed by statistics (mean and std. dev.) pooling layer which converts frame level features to utterance level features.  These segment level statistics are fed through two layers of feed forward network with $512$ hidden dimensions. The x-vectors are the embeddings from the affine layer after the statistics pooling (output of layer $6$ before the non-linearity with $512$ dimensions). The target for the x-vector model are the language labels. We use the LRE2017 training data along with data augmentation to train the x-vector model.
\textcolor{black}{The model used a five fold data augmentation procedure with  room reverberation as well as the additive noises like music, noise and babble (MUSAN corpus~\cite{snyder2015musan}) at SNR values of $5,10,15,20$ dB. 
}
The entire model is trained using the Kaldi recipe~\cite{povey2011kaldi}. We report the results using x-vector features for the LRE2017 evaluation in Table~\ref{table:lre17_perf}. \textcolor{black}{We also report the performance of the x-vector system directly using the posteriors from the trained network (denoted as x-post).}}

\begin{table*}[t!]
\centering
\caption{\textcolor{black}{Performance of the systems with partial noise ($50$\% of the utterance, $10$ sec. recording condition)} at various SNR levels. \textcolor{black}{The x-vector based systems are shown in parentheses.} }
\label{table:part_noisy}
\begin{tabular}{|c|c|c|c|c|c|c|c|}
\hline

SNR & i-LDA-SVM \textcolor{black}{(x-LDA-SVM)} & LSTM \cite{2016_odyssey_lstm_nist_Zazo}  & DNN-Attn \cite{vuddagiri2018curriculum} &  i-BLSTM  \textcolor{black}{(x-BLSTM)} & HGRU & \textcolor{black}{(x-post)}  & \textcolor{black}{(x-BLSTM-E2E)} \\ \hline
\multicolumn{7}{|c|}{Accuracy (\%)}                           \\ \hline

5dB        & 53.16 \textcolor{black}{(57.36)}   & 56.50 & 54.64    &  {59.79} \textcolor{black}{(61.41)} & 55.32 & \textcolor{black}{({58.85})} & \textcolor{black}{(\textbf{61.68})} \\ 
10dB       & 55.57 \textcolor{black}{(61.07)}  & 60.42 & 58.22    &  {63.02} \textcolor{black}{(63.56)}& 60.20 & \textcolor{black}{({61.44})} & \textcolor{black}{(\textbf{63.64})}\\ 
15dB       & 58.33 \textcolor{black}{(64.10)}  & 62.61 & 61.61    &  \textbf{65.90} \textcolor{black}{(65.66)} & 63.16 & \textcolor{black}{({63.05})}& \textcolor{black}{(65.60)} \\ 
20dB       & 59.56 \textcolor{black}{(65.53)}  & 64.61 & 63.55    & \textbf{68.16} \textcolor{black}{(66.51)} & 65.65 & \textcolor{black}{({63.79})} &  \textcolor{black}{(66.16)}\\ \hline
Avg. & 56.65 \textcolor{black}{(62.01)} &61.04 &59.50 &{64.20} \textcolor{black}{(\textbf{64.29})} & 61.08 & \textcolor{black}{({61.78})} & \textcolor{black}{(64.27)} \\ \hline \hline
\multicolumn{6}{|c|}{Cavg}                           \\ \hline 
5dB       & 0.54 \textcolor{black}{(0.66)} & 0.52 & 0.50 & {0.47} \textcolor{black}{(\textbf{0.38})} & 0.57 & \textcolor{black}{(0.45)}& \textcolor{black}{(0.44)} \\ 
10dB      & 0.50 \textcolor{black}{(0.61)} & 0.47 & 0.44 & {0.43} \textcolor{black}{(\textbf{0.35})} & 0.51 & \textcolor{black}{(0.41)}& \textcolor{black}{(0.41)} \\ 
15dB      & 0.47 \textcolor{black}{(0.58)} & 0.44 & 0.41 & {0.39} \textcolor{black}{(\textbf{0.33})} & 0.47 & \textcolor{black}{(0.38)}& \textcolor{black}{(0.38)} \\ 
20dB      & 0.45 \textcolor{black}{(0.56)} & 0.42 & 0.39 & {0.37} \textcolor{black}{(\textbf{0.32})} & 0.44 & \textcolor{black}{(0.37)} & \textcolor{black}{(0.37)}\\ \hline
Avg. & 0.49 & 0.46 & 0.44 & {0.42} \textcolor{black}{(\textbf{0.35})} & 0.50 & \textcolor{black}{(0.40)} & \textcolor{black}{(0.40)}\\ \hline \hline
\multicolumn{6}{|c|}{EER (\%)}                           \\ \hline
5dB       & 17.56 \textcolor{black}{(22.32)} & 14.59 & 13.28 & {12.30} \textcolor{black}{(\textbf{10.67})} & 15.36 & \textcolor{black}{(13.81)}& \textcolor{black}{(12.54)} \\ 
10dB      & 15.79 \textcolor{black}{(20.66)} & 13.10 & 11.52 & {10.84} \textcolor{black}{(\textbf{9.75})} & 13.05 & \textcolor{black}{(12.41)} & \textcolor{black}{(11.22)} \\ 
15dB      & 14.32 \textcolor{black}{(18.75)} & 11.80 & 10.45 & {9.60} \textcolor{black}{(\textbf{9.04})} & 11.89 & \textcolor{black}{(11.60)} & \textcolor{black}{(10.62)}  \\ 
20dB      & 13.70 \textcolor{black}{(17.97)} & 11.20 & 9.64 & {8.95} \textcolor{black}{(\textbf{8.59})} & 11.08 & \textcolor{black}{(11.34)} & \textcolor{black}{(10.22)} \\ \hline
Avg. & 15.34 \textcolor{black}{(19.92)} & 12.67 &11.22 & {10.42} \textcolor{black}{(\textbf{9.51})} & 12.85 & \textcolor{black}{(12.29)} & \textcolor{black}{(11.15)} \\ \hline
\end{tabular}
\end{table*}
\subsection{DNN and LSTM baseline}
We implement the  best performing LSTM model \cite{2016_odyssey_lstm_nist_Zazo}, which is a two layer LSTM with $512$ units in each layer followed by an output softmax layer as a baseline end to end system. 
We also compare the performance of proposed models with a DNN with attention developed recently \cite{vuddagiri2018curriculum}. 

\subsection{Proposed Relevance Models for LID}
The i-BLSTM model uses the  i-vector extraction setup similar to the one used in the baseline system. The audio recordings are chunked into $1000$ msec. which are shifted every $200$ msec. The segment i-vectors are used to train the BLSTM model  on the training data. We use $80,000$ recordings of $15$ sec. duration  from the LRE training dataset and the model is trained using a cross entropy loss.  For training the proposed end-to-end HGRU model on the LRE17 data, audio snippets of duration ranging from $3$ sec. to $30$ sec. are randomly sampled from the training data. A similar kind of setup is used for RATS dataset, where around $100,000$ samples of $15$ sec. duration is sampled for training the i-BLSTM model and random sampling of $3$ sec. and $30$ sec. is done for the end-to-end HGRU model.
\textcolor{black}{Finally, we also experiment with replacing the i-vector features with the x-vector features for the BLSTM model (this model is refered to as x-BLSTM system).}

\section{Language Recognition Results}\label{sec:results}
\subsection{LRE Evaluation Results}
The results for LRE evaluations are listed in Table~\ref{table:lre17_perf}. 
The relevance weighted models proposed in this paper are the relevance weighted Baum-Welch statistics (RWBW), the hybrid i-vector BLSTM model (i-BLSTM) and the hierarchical GRU model (HGRU). 

As seen in Table~\ref{table:lre17_perf}, in terms of the $C_{avg}$, the baseline neural models like LSTM \cite{2016_odyssey_lstm_nist_Zazo} and DNN with attention \cite{vuddagiri2018curriculum} perform comparatively well on short durations ($3$ sec. and $10$ sec.). However, on longer duration of $30$ sec. and $1000$ sec, the i-vector based LDA-SVM system provides the best performance. This indicates that the limitations of the previous models to incorporate the long term dependencies.
The proposed relevance based approaches compare favorably with the i-vector SVM baseline in all conditions. The RWBW system provides the best performance on the long duration ($1000$ sec.) condition as there are a large number of short segment i-vectors in these recordings. \textcolor{black}{However, in short duration conditions, the RWBW is inferior to other proposed methods.} The HGRU model performs similar to the DNN-attn model proposed previously. The best system in terms of the $C_{avg}$. and the EER metric is the x-BLSTM system. 

\begin{figure}[t!]
    \centering
    \includegraphics[width=8cm]{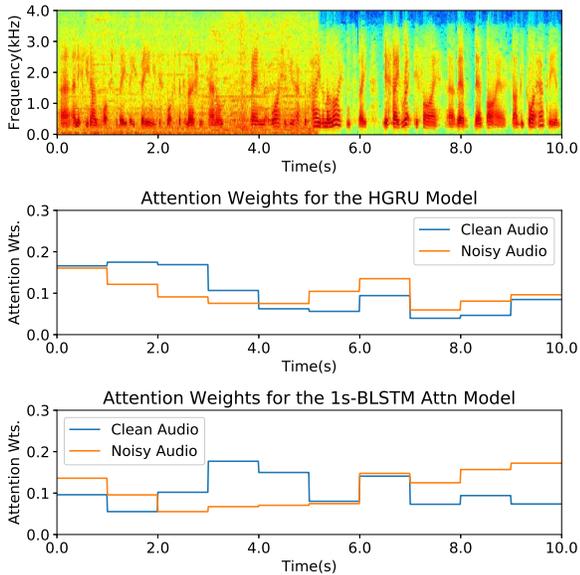}
    \caption{Plot of a partially noisy (noise present in the first half of the audio) LRE-17 audio file spectrogram and the corresponding attention weights estimated every $1000$ msec. for the proposed HGRU model and the i-BLSTM model. 
    }
    \label{fig:noisy_lre_attn}
    \vspace{-0.3cm}
\end{figure}
\begin{figure*}[t!]
     \centering
     \includegraphics[width=15cm,height=10cm]{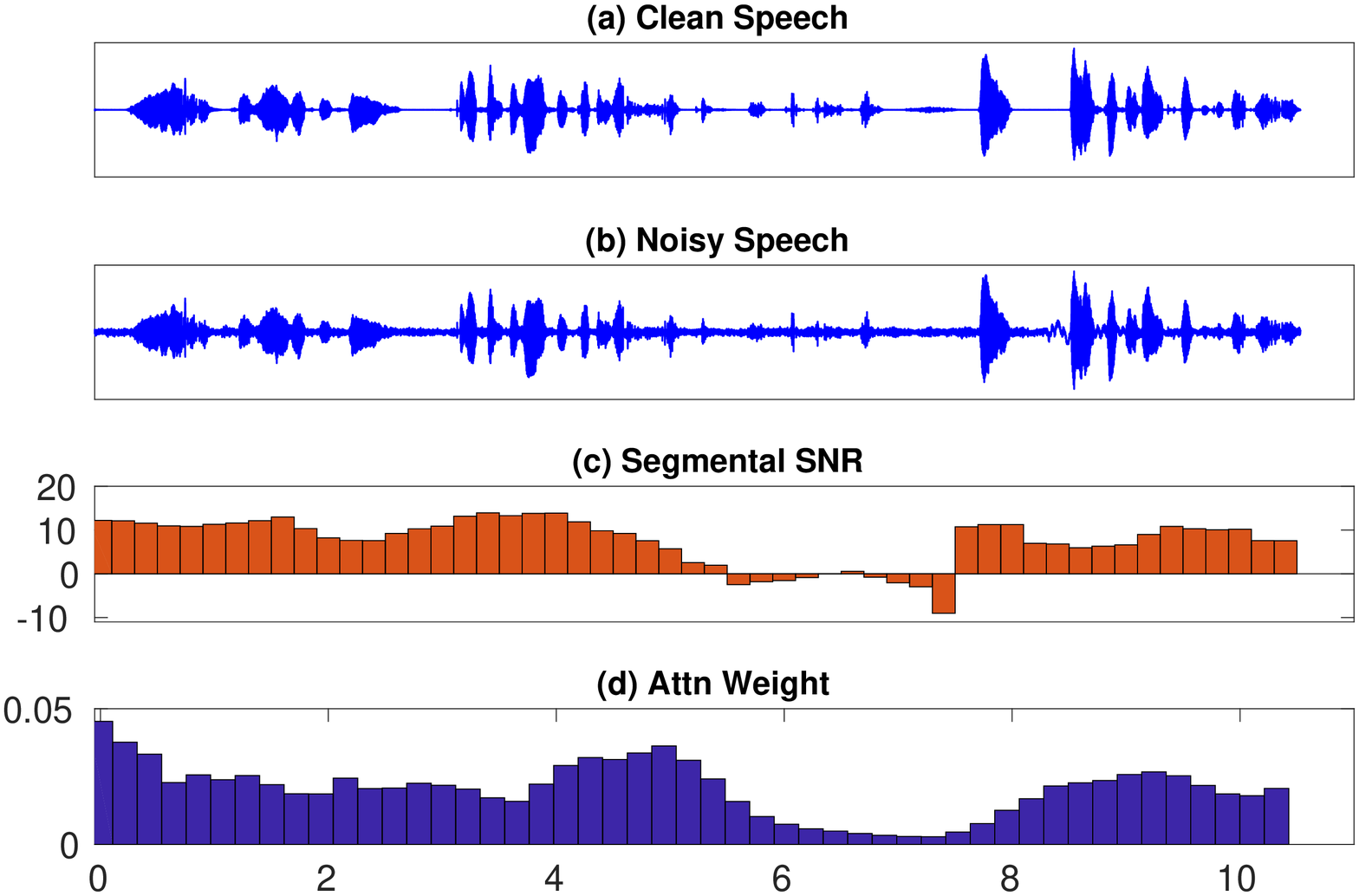}
     \caption{\textcolor{black}{Plot of (a) a clean LRE utterance from validation data, (b) the same utterance with $10$ dB babble noise, (c) segment level SNR measured with $1000$msec. windows shifted by $200$msec. and (d) the corresponding attention weights estimated from the i-BLSTM model for $1000$ msec. windows shifted by $200$ msec. }}
     \label{fig:anr_lre_attn}
     \vspace{-0.3cm}
 \end{figure*}
\subsection{Results on Noisy LRE Data}
The results with various noisy versions of the evaluation data are shown in Table~\ref{table:noisy} and Table~\ref{table:part_noisy}. 
For the experiments with noisy data (Table~\ref{table:noisy}), among the various baseline systems, the LSTM model \cite{2016_odyssey_lstm_nist_Zazo} gives the best robustness.  The proposed approach of hybrid i-BLSTM model provides significant improvements over all models considered  for all SNR conditions. The hierarchical GRU model also improves over the LDA-SVM baseline but the performance of this system is inferior to the hybrid x-BLSTM system as well as the DNN-attn model \cite{vuddagiri2018curriculum}.  On the average, the proposed x-BLSTM model shows a relative improvement of $28.8$ \% over the baseline i-LDA-SVM in terms of the accuracy measure. 

On the partially noisy conditions (Table~\ref{table:part_noisy}), the trends are similar. The baseline performance of the LDA-SVM system is improved by the DNN-attn model. However, the proposed i-BLSTM model improves over the previous neural network model based approaches for LID. On the average, the x-BLSTM model improves the baseline LDA-SVM relatively by $28.5$\% in term of the accuracy measure. 

\textcolor{black}{The reason for the counterintuitive lack of improvement for the partial noise over the full noise case  for the baseline systems (comparing Table~\ref{table:noisy} and Table~\ref{table:part_noisy} for x-post system results) may be attributed to the lack of ability of the baseline methods like x-post in modeling the non-stationarity of the signal (partial noise). The results for the proposed approach (x-BLSTM) in terms of [Acc]\{$C_avg$\}(EER) reported in Table III are [63.15]\{0.37\}(10.80). These results are improved for the partial noise case in  Table IV as [64.29]\{0.35\}(9.51). We attribute this improvement to the ability of the attention modeling to re-weight the relevance to regions of the signal that are less noisy.} 

\begin{table}[t!]
\centering
\caption{Performance of the baseline system and the proposed models on the RATS dataset in terms of  accuracy, $C_{avg}$ and EER. }
\label{table:rats_perf}
\begin{tabular}{|c|c||c|c|}
\hline
Duration (sec.) & i-LDA-SVM & i-BLSTM & HGRU  \\ \hline
\multicolumn{4}{|c|}{ Accuracy (\%)}                                      \\ \hline
3               &   65.39   &   64.16   &  \textbf{69.78}     \\ 
10              &   77.46   &   79.07   &    \textbf{81.76}   \\ 
30              &   85.38   &   87.12  &  \textbf{87.89}    \\ 
120     &   92.22   &   \textbf{92.69}   &  91.43     \\ \hline
\multicolumn{4}{|c|}{Cavg}                                             \\ \hline
3               & 1.12   &   1.04   & \textbf{0.84}  \\ 
10              & 0.81   &   0.63   & \textbf{0.57}  \\ 
30              & 0.57   &   \textbf{0.43}   & \textbf{0.43}  \\ 
120    & 0.40   &   \textbf{0.32}   & 0.35  \\ \hline
\multicolumn{4}{|c|}{EER (\%)}                  \\ \hline
3               &   25.90   &   24.32    &  \textbf{19.56}     \\ 
10              &   17.81   &  13.69   &    \textbf{12.23}     \\ 
30              &   11.88   &   \textbf{8.61}   &    8.87  \\
120    &   7.64   &  \textbf{5.74}   &  6.81  \\ \hline
\end{tabular}
\end{table}

\subsection{Results on RATS Dataset}
 Here, we compare the performance of the i-vector LDA-SVM with the proposed approaches of hybrid i-BLSTM system and the end to end HGRU model. As seen in the results, the proposed approaches  yields significant improvements over the baseline system in terms of all the performance metrics and for all the  duration conditions.  The HGRU model is more efficient in the short duration conditions while the i-BLSTM model improves over the HGRU model for longer duration, 
 full length ($120$ sec.) recording conditions. 

\subsection{\textcolor{black}{Results with x-vector embeddings} } 
\textcolor{black}{From Table~\ref{table:lre17_perf}, 
we observe that the x-vector based LDA-SVM improves over the baseline i-vector based system with similar back-end in terms of $C_{avg}$ measure for all durations. The proposed BLSTM based backend improves the x-LDA-SVM system significantly for short duration ($3$sec. condition). However, for the longer durations, the baseline xvec. LDA-SVM provides better performance than the proposed model. \textcolor{black}{The x-vector network posteriors directly used for LRE is also worse compared to the use of the SVM backend (especially for long durations).}}

\textcolor{black}{The results on noisy LRE17 dataset and partially noisy dataset are given in  Table~\ref{table:noisy} and Table~\ref{table:part_noisy} respectively. As seen here, the x-vector embeddings provide noticeable improvements over the i-vector counterparts. Further, the proposed backend approach of x-BLSTM provides significant improvements over the baseline system both in the noisy and partially noisy case. The relative improvements in terms of $C_{avg}$ metric over the baseline system ranges from $34-43$ \% for noisy conditions and about $42$ \% for partially noisy conditions. It is also noteworthy that the proposed approach suffers from a performance degradation of less than $20$\% relative (in terms of $C_{avg}$) compared with clean conditions when then SNR is above $15$ dB while the baseline x-LDA-SVM system suffers from more than $40$ \% degradation in relative performance. These results are also in alignment with i-vector based results.
}

\section{Discussion and Analysis}\label{sec:discussion}

\subsection{Importance of Relevance Modeling}
The plot shown in Fig.~\ref{fig:noisy_lre_attn} illustrates the spectrogram of a partially noisy (noise present in the first half of the audio recording) and the corresponding attention weights from the HGRU and the i-BLSTM model. The attention weights for the same audio recordings without any noise (clean) condition are also shown for reference. As seen in this plot, the attention weights are considerably changed due to the presence of the noise in the first half of the file. The model is able to be adaptive and focuses the language detection on the more reliable regions of the audio.  
The increased senstivity of the i-BLSTM system also potentially explains the improved performance of the i-BLSTM model in the experiments reported in Table~\ref{table:noisy} and Table~\ref{table:part_noisy}.

\textcolor{black}{Specifically, in the presence of noise, different parts of the signal have different signal-to-noise ratio values (as the speech signal is time-varying even if the noise is stationary). This can be visualized using the example given in Fig.~\ref{fig:anr_lre_attn}. While the SNR for this entire speech utterance is $10$ dB, the SNR measured at short-term segment level is highly time varying (ranging between $20$ dB and $-10$ dB). 
The goal of the relevance models for LID proposed in this work is to deemphasize the contributions from the noisy regions while enhancing the contribution of the high SNR regions in making the decision about the language identity. An example of the attention weights from the i-BLSTM model is also shown in Fig.~\ref{fig:anr_lre_attn} (d). As seen in this plot, the attention weights tend to track the short-term SNR thereby weighting the embeddings from the regions of high SNR with higher weights while reducing the weights associated with regions of low SNR.}

\subsection{\textcolor{black}{End-to-end LID with x-vector}}
\textcolor{black}{With use of the x-vector based embeddings in the proposed approach, we explore whether the proposed attention based BLSTM and the embedding x-vector model can be jointly trained. This model is referred to as the x-BLSTM E2E. The results for clean LRE17 test set are also reported in Table~\ref{table:lre17_perf} and the noisy LRE results are reported in Table~\ref{table:noisy} and Table~\ref{table:part_noisy}. As seen here, the end-to-end system improves over the x-BLSTM system for short durations while the performance is degraded for longer duration conditions. \textcolor{black}{It is also important to note that the performance of the x-BLSTM-E2E is better than the x-post system which uses the x-vector network outputs directly for LRE. These results  indicate that the attention based backend modeling is important for noisy LRE task.} 
}

\section{Conclusion}\label{sec:summary}
In this paper, we have proposed a new hybrid i-vector/x-vector BLSTM attention model (i/x-BLSTM) for language recognition where the sequence of $1000$ msec. i-vectors/x-vectors are modeled in a bidirectional attention based network.  A novel processing pipeline with hierarchical gated recurrent unit (HGRU) or x-vector BLSTM model is also proposed for end-to-end spoken language recognition. While, the conventional backend systems based on LDA-SVM classifiers ignore sequential speech information, the attention mechanism in the proposed model plays the role of relevance weighting, where portions of the speech signal more relevant to classification decision are given higher weightage. This relevance modeling is shown in particular to improve the robustness of the model for language recognition in noisy conditions. With several experiments on the noisy LRE dataset as well as the RATS dataset, we show the effectiveness of the proposed model. We also present a detailed analysis, which highlights the role of attention modeling in language recognition. 



%

\section*{Acknowledgment}
The authors would like to acknowledge Shreyas Ramoji, Satish Kumar and Vaishnavi Y for their help in setting up the LRE 2017 baseline, Omid Sadjadi for code fragments implementing the NIST LRE baseline system.
 The work reported in this paper was partly funded by grants from the Department of Science and Technology (EMR/2016/007934) and by Department of Atomic Energy (DAE/34/20/12/2018-BRNS/34088). 
\ifCLASSOPTIONcaptionsoff
  \newpage
\fi



%
\bibliographystyle{IEEEbib}
\bibliography{references}
\vspace{-1cm}
%
\begin{IEEEbiography}[{\includegraphics[width=1in,height=1.25in,clip,keepaspectratio]{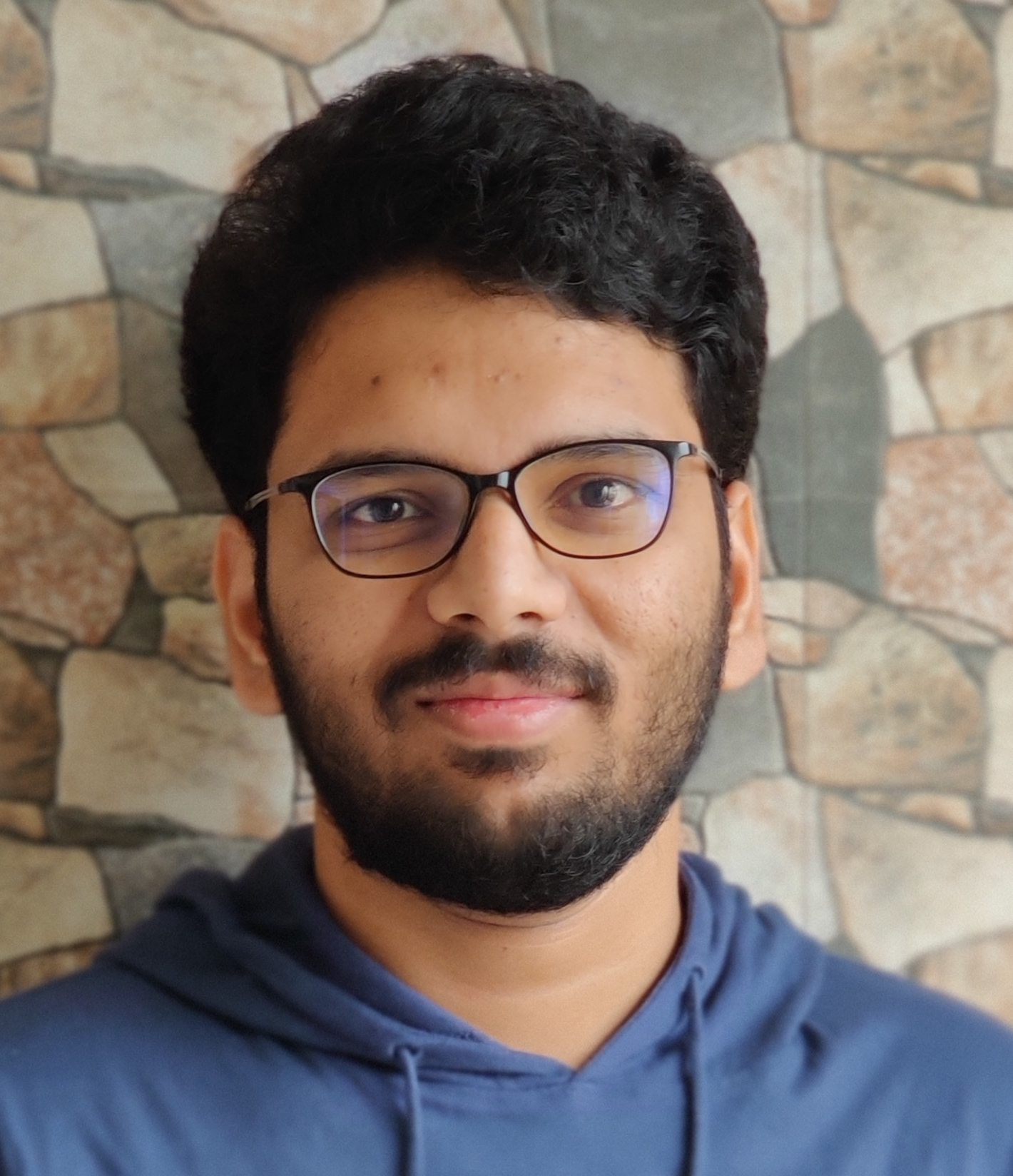}}]{Bharat Padi}
 received his Bachelor of Engineering in Electrical and Electronics Engineering from Andhra
University, Visakhapatnam, India in 2009 and his Master's Degree in System Science and
Automation from Indian Institute of Science (IISc), Bengaluru, India in 2018. Between
2009-2016, he worked as a systems engineer in Infosys Technologies and later as an
Assistant Manager (Electrical) in Visakhapatnam steel plant.
He is currently working in minds.ai as a neural network research engineer aiming at
developing controllers for vehicles using deep reinforcement learning.
His current research interests include developing deep end-end models for spoken language recognition and reinforcement learning. He is a member of the IEEE. 
\end{IEEEbiography}
\vspace{-1cm}
\begin{IEEEbiography}[{\includegraphics[width=1in,height=1.25in,clip,keepaspectratio]{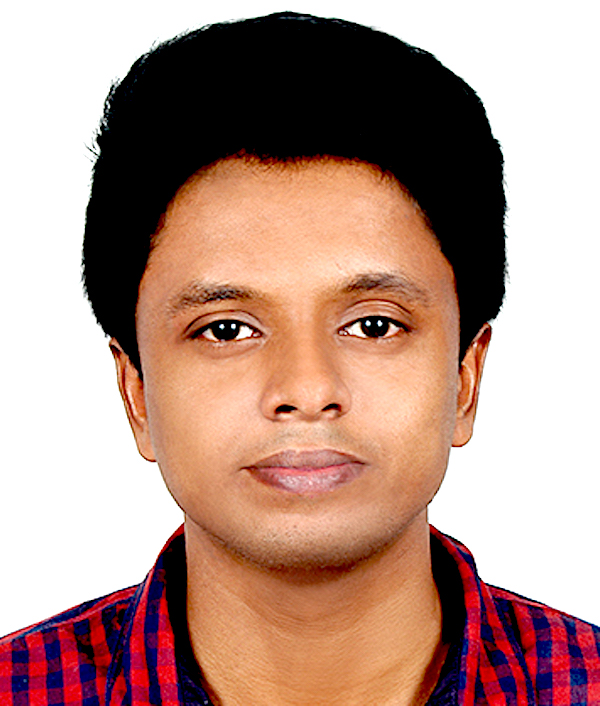}}]{Anand Mohan}
 is currently working as Applied Scientist in Alexa Speech team in Amazon, Bangalore, India. He received his Master's Degree in Artificial Intelligence from Indian Institute of Science (IISc), Bangalore, India in 2019 and his Bachelor's Degree in Electronics and Communication Engineering from National Institute of Technology, Calicut, India in 2015. His research interests include end-to-end ASR technologies, speaker and language identification. 
\end{IEEEbiography}

\begin{IEEEbiography}[{\includegraphics[width=1in,height=1.25in,clip,keepaspectratio]{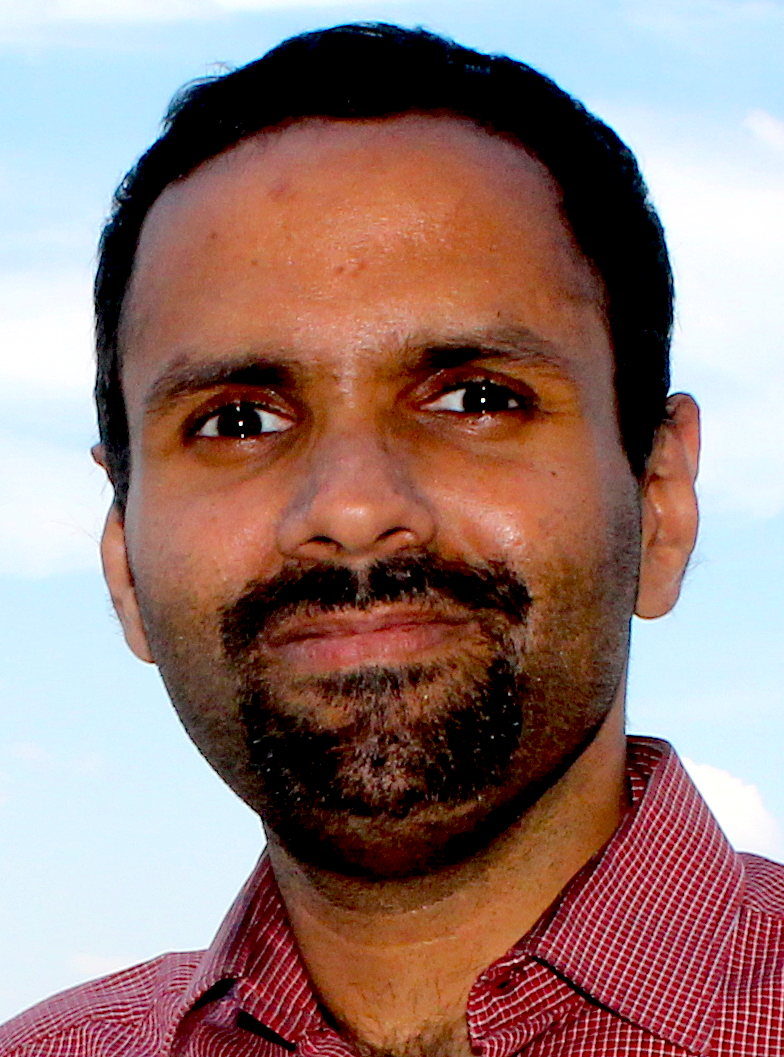}}]{Sriram Ganapathy}
is a faculty member at the Electrical Engineering, Indian Institute of Science, Bangalore, where he heads the activities of the learning and extraction of acoustic patterns (LEAP) lab. Prior to joining the Indian Institute of Science in 2016, he was a research staff member at the IBM Watson Research Center, Yorktown Heights. He received his Doctor of Philosophy from the Center for Language and Speech Processing, Johns Hopkins University in 2012. He obtained his Bachelor of Technology from College of Engineering, Trivandrum, India in 2004 and Master of Engineering from the Indian Institute of Science, Bangalore in 2006. He has also worked as a Research Assistant in Idiap Research Institute, Switzerland from 2006 to 2008.  

At the LEAP lab,  his research interests include signal processing, machine learning and deep learning methodologies for speech/speaker recognition and auditory neuroscience. He is a subject editor for the Speech Communications journal, member of ISCA and a senior member of the IEEE. 

\end{IEEEbiography}








\end{document}